\documentclass[]{spie}

\usepackage{amsmath,amsfonts,amssymb}
\usepackage{graphicx}
\usepackage[colorlinks=true, allcolors=blue]{hyperref}

\title{Installation and first commissioning results of the JEF lead tungstate calorimeter}

\author[a]{Alexander Somov}
\author[a]{Vladimir Berdnikov}
\affil[a]{Thomas Jefferson National Accelerator Facility, Newport News, Virginia 23606, USA}

\authorinfo{Further author information: (Send correspondence to A.Somov)\\A.Somov: E-mail: somov@jlab.org.edu, Telephone: 1 757 269 5553\\  V.Berdnikov: E-mail: berdnik@jlab.org, Telephone: 1 757 269 6928\\
for the JEF working group and GlueX collaboration}

\begin{document} 
\maketitle

\begin{abstract}
 The Electromagnetic Calorimeter (ECAL), consisting of 1,596 lead tungstate scintillating crystals, has been recently constructed and 
installed in Experimental Hall D at Jefferson Lab (JLab). The calorimeter is a key component of the JLab Eta Factory Experiment, whose 
main goal is to measure the decays of $\eta$ and $\eta^{\prime}$ mesons into multi-photon final states. The ECAL replaces the inner part 
of the former forward lead-glass calorimeter. Scintillation light from each crystal is detected using Hamamatsu R4125 photomultiplier tubes. 
Calorimeter modules were fabricated and tested in the lab using light from light-emitting diodes before being installed in the detector frame. 
The detector is currently undergoing commissioning using the light monitoring system and cosmic rays. We will present an overview of the 
fabrication and testing of the calorimete  modules, along with the first detector commissioning results. The ECAL was ready for data-taking in spring 2025.
\end{abstract}


\section{INTRODUCTION}
\label{sec:intro}  

The GlueX detector~\cite{gluex} in experimental Hall D at Jefferson Lab is a forward magnetic spectrometer, which is used to carry 
out experiments using a beam of photons. The main goal of the Jefferson Lab Eta Factory (JEF) experiment~\cite{jef,somov_jef}  is 
to measure the decays of $\eta$ and $\eta^{\prime}$ mesons into multi-photon final states. The JEF experiment runs in parallel with 
the GlueX II experiment~\cite{gluex_proposal} and began taking data in April 2025. To enable good reconstruction of photons from $\eta$ 
meson decays required by the JEF physics program, we constructed an electromagnetic calorimeter composed of 1,596 lead tungstate (PbWO$_4$) 
scintillating crystals. The ECAL replaced the inner part of the existing GlueX forward lead glass calorimeter~\cite{gluex}, which originally 
consisted of 2800 modules. The lead tungstate material offers several advantages over lead glass, including a smaller radiation length, 
and Moli$\grave{\rm e}$re radius, higher light yield, and superior radiation hardness. These properties result in approximately a factor 
of two improvement in energy resolution, expected to be better than $\Delta E / E \sim 3\%$, and a factor of four enhancement in detector 
granularity compared to the previous detector. The total cost of the calorimeter upgrade was approximately $\$5$ million. Construction of 
the ECAL modules began in 2022 and was completed in about three years. The new calorimeter was fully integrated into the GlueX detector’s 
trigger and data acquisition systems and was successfully commissioned in April 2025. Twenty-eight summer students from 11 universities 
participated in the project. This article is organized as follows: Section 2 describes the assembly and testing of the ECAL modules; 
Section 3 presents the detector installation and its integration into the GlueX infrastructure; and Section 4 discusses the initial commissioning 
results of the new detector.

\begin{figure} [ht]
\begin{center}
\begin{tabular}{c} 
\includegraphics[height=5.5cm]{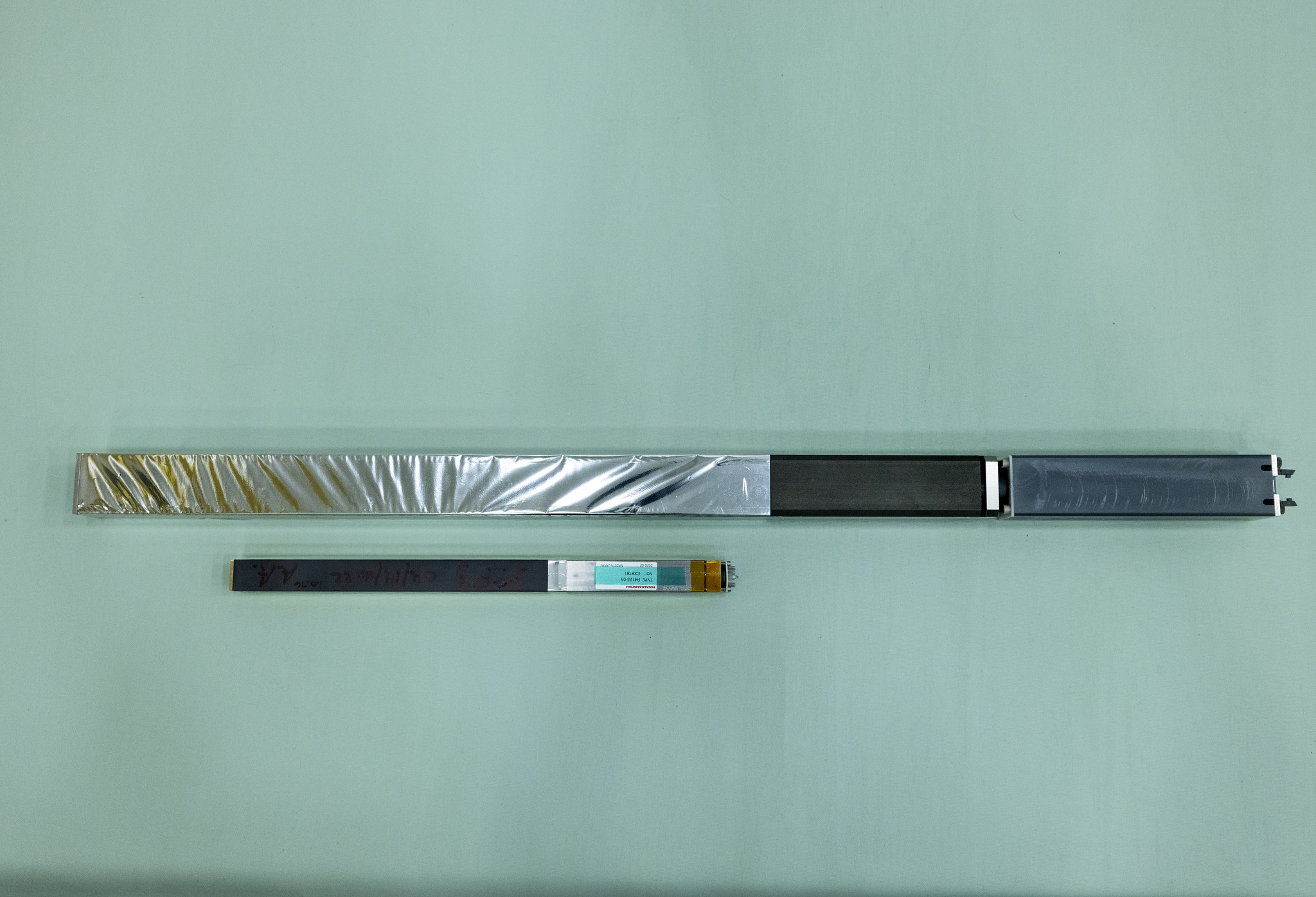}
\includegraphics[height=5.5cm]{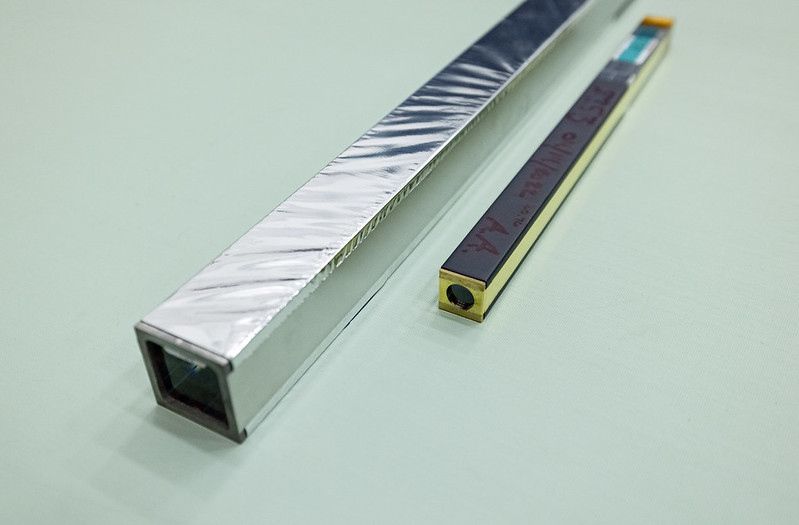}
\end{tabular}
\end{center}
\caption[example] 
{ \label{fig:ecal_module} 
Lead glass and lead tungstate modules.}
\end{figure} 
\begin{figure} [ht]
\begin{center}
\begin{tabular}{c} 
\includegraphics[height=6cm]{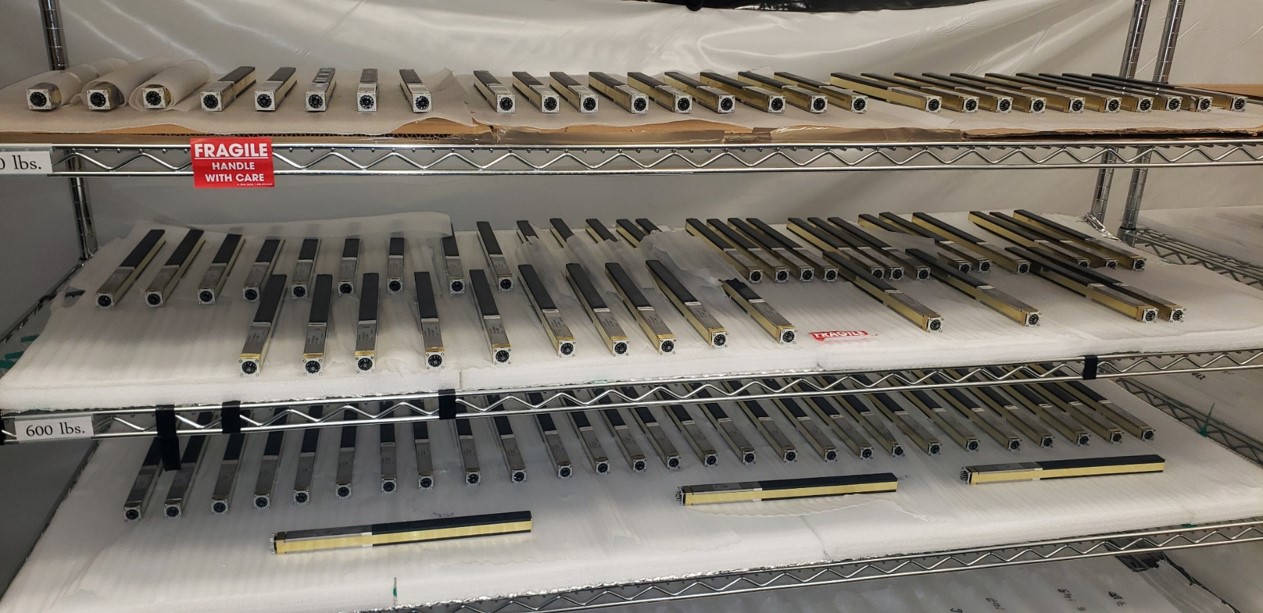}
\end{tabular}
\end{center}
\caption[example] 
{ \label{fig:ecal_assembled} 
Assembled ECAL modules.}
\end{figure} 

\section{ECAL MODULE ASSEMBLING AND TESTING}

The calorimeter features a modular design~\cite{somov_ecal,ecal_nim}. A total of 1596 modules form a $40\times40$ array, with a central of $2\times2$ hole for the photon beam. Each lead tungstate crystal, measuring 2.05 cm × 2.05 cm × 20 cm, is wrapped in Enhanced Specular Reflector foil (produced by ${\rm 3M}^{\rm TM}$) and light-tight Tedlar. Scintillation light is collected using a Hamamatsu R4125 photomultiplier tube (PMT), which is wrapped with mu-metal foil and placed inside an Amumetal cylinder and soft iron housing to reduce the effects of stray magnetic fields produced by the GlueX solenoid. The PMT is coupled to the crystal via an acrylic light guide, which extends the magnetic shielding beyond the photocathode face of the PMT. One end of the light guide is glued to the PMT, while the other is attached to the crystal using optical silicone rubber. The ECAL module is shown Fig.~\ref{fig:ecal_module} along with the lead glass module. Due to the significantly shorter radiation length of PbWO$_4$, the ECAL enables a much more compact detector design and better detector granularity.

The project began with the characterization and quality control of lead tungstate crystals, which were procured from two vendors: the Shanghai Institute of Ceramics (SICCAS) and CRYTUR in the Czech Republic. This phase was followed by the preparation of components required for module assembly, including gluing photomultiplier tubes  to light guides, pre-shaping reflective foils, producing silicone optical pads ("cookies"), and other assembly tasks. Module fabrication commenced in early 2022 and was finished by middle 2023. Approximately 1,620 modules were assembled at Jefferson Lab. A set of assembled ECAL modules is presented in Fig.~\ref{fig:ecal_assembled}. 

The original Hamamatsu PMT divider, which supplies voltages to the PMT dynodes and reads out signal pulses, was modified at Jefferson Lab by adding two transistors to the divider chain and an amplifier with a gain of approximately three~\cite{divider}. The amplifier was placed directly on the divider's printed circuit board. These modifications allowed reducing operating high voltage and the anode current, thereby improving performance under high-rate conditions. The amplifier requires an additional power supply of $\pm 5{\rm V}$. The modified PMT divider is referred to as the PMT active base. The active bases were attached to the ECAL modules after their installation in the experimental hall. Signal pulses from the PMTs are digitized using an analog-to-digital converter (ADC) operating at a sampling rate of 250 MHz.

\begin{figure} [t]
\begin{center}
\begin{tabular}{c} 
\includegraphics[height=4.6cm]{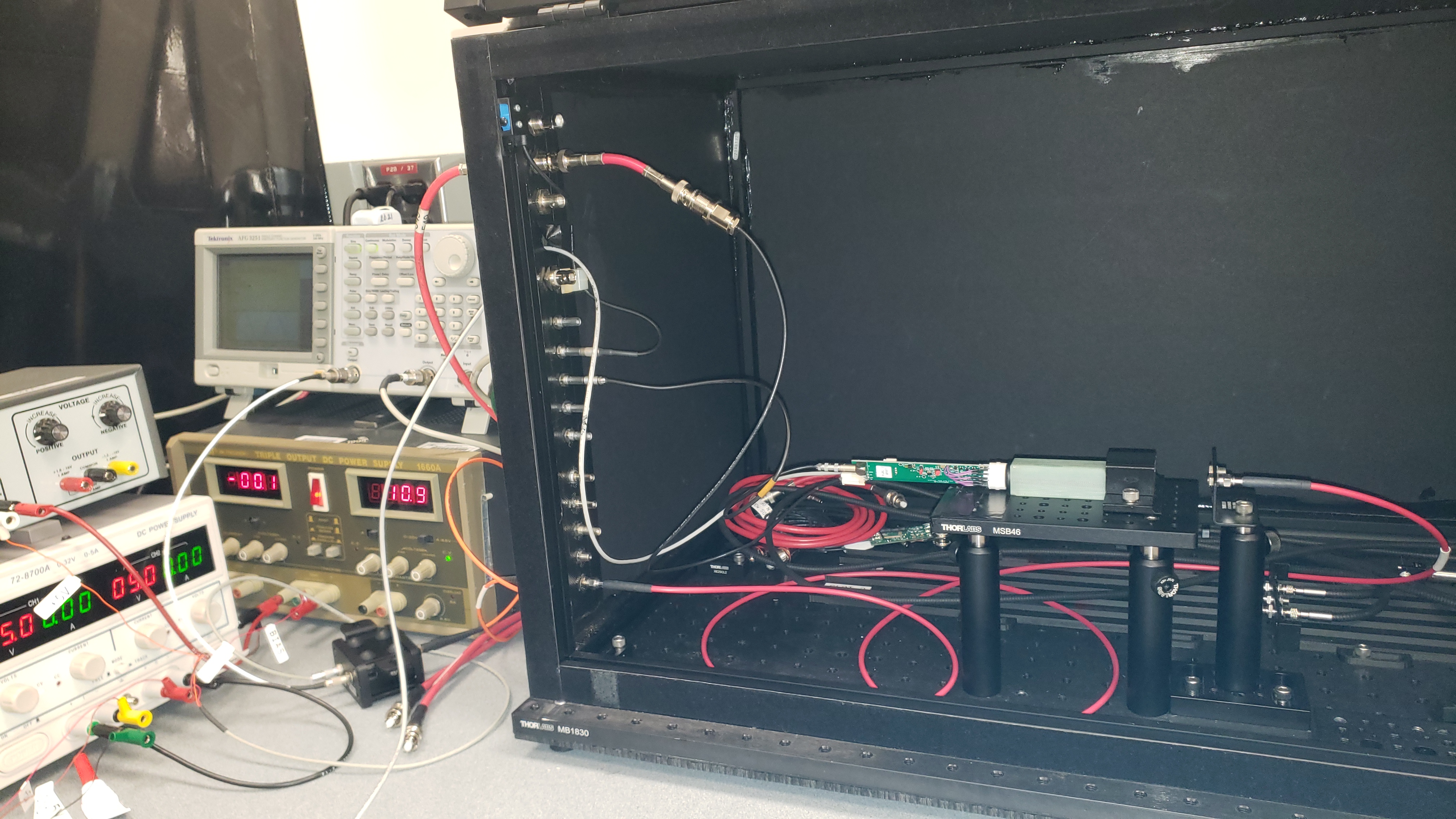}
\hspace{0.05in}
\includegraphics[height=4.8cm]{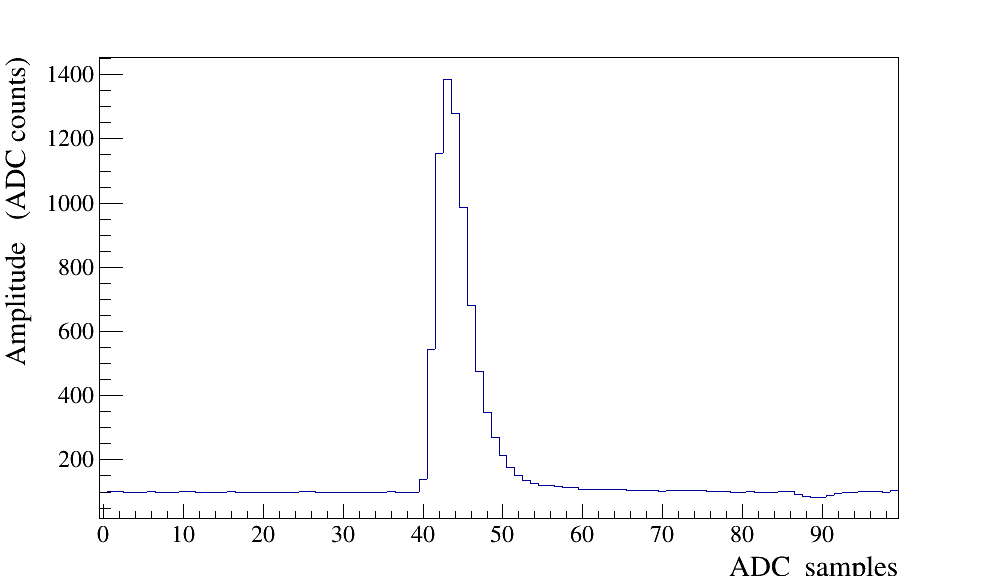}
\end{tabular}
\end{center}
\caption[example] 
{ \label{fig:testing_dividers} 
Setup for testing ECAL dividers and assembled modules using the light monitoring system. The signal waveform induced by the LMS is digitized by a flash ADC (right). The flash ADC operates at a sampling rate of 250 MHz, corresponding to a sample interval of 4 ns.}
\end{figure} 

All assembled ECAL modules were tested using a pulser system based on a blue light-emitting diode (LED). Each module was placed inside a light-tight box, and light from the LED was delivered to the front face of the module via an optical fiber. The test setup is shown in Fig.~\ref{fig:testing_dividers}. The signal waveform and pulse amplitude produced by the LED were examined using an oscilloscope. The same setup was also used to test the PMT active bases. In this case, each base was connected to a reference PMT operated at a fixed high voltage of 1 kV. Both the pulse amplitude and the current drawn by the active base were measured, with the latter being approximately 400~$\mu$A.

\section{DETECTOR INSTALLATION}

\begin{figure} [b]
\begin{center}
\begin{tabular}{c} 
\includegraphics[height=8cm]{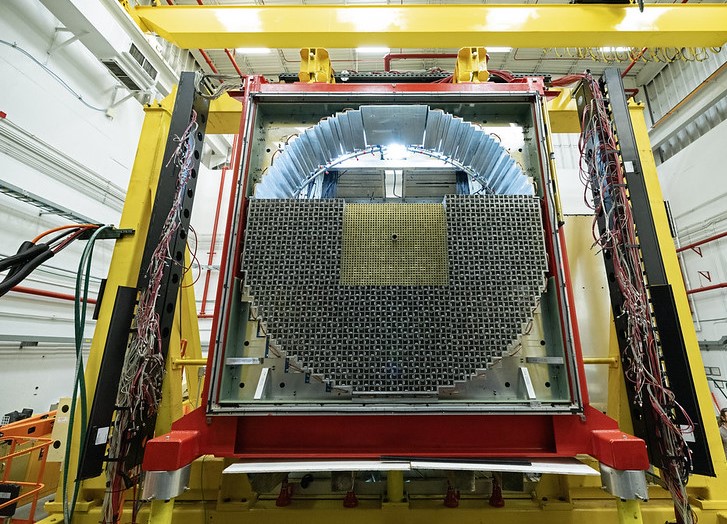}
\end{tabular}
\end{center}
\caption[example] 
{ \label{fig:ecal_installation} 
Installing ECAL in experimental Hall D. The ECAL modules at the center are surrounded by lead-glass modules.}
\end{figure}

Installation of the ECAL in the experimental hall began in April 2023 with the removal of 2,800 lead glass modules from the detector frame. A new frame, equipped with water-based cooling blocks, was designed and installed by the GlueX technical team. The cooling system is supplied by two chillers. Maintaining a constant temperature is essential for the detector, as the light yield of the PbWO$_4$ crystals is temperature-dependent. The lead glass modules located beneath the ECAL were installed first, followed by the stacking of the ECAL modules. The installation process is shown in Fig.~\ref{fig:ecal_installation}.

\begin{figure} [t]
\begin{center}
\begin{tabular}{c} 
\includegraphics[height=6cm]{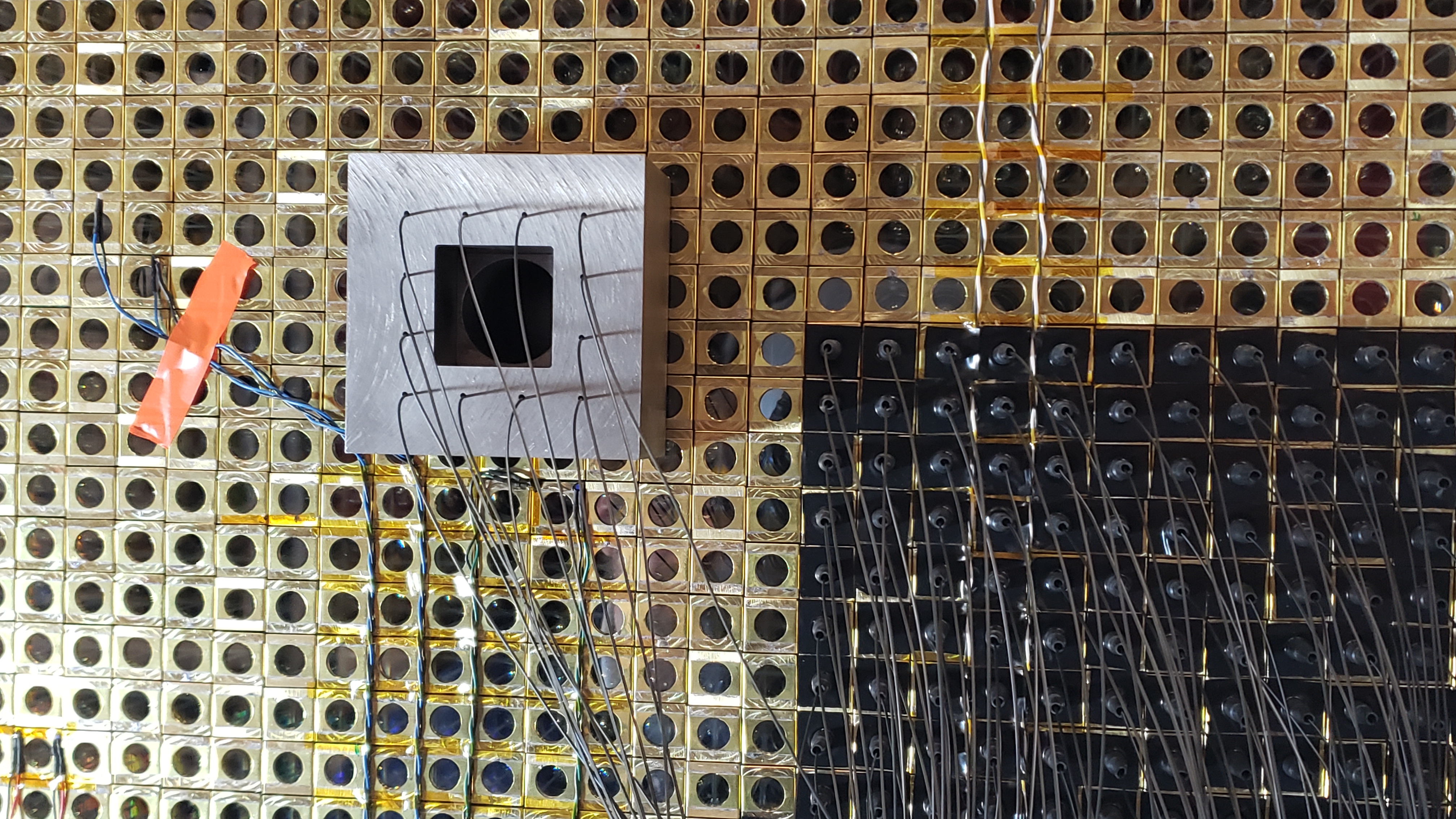}
\end{tabular}
\end{center}
\caption[example] 
{ \label{fig:install_fibers} 
Installation of optical fibers for the light monitoring system. Each individual fiber is glued to the crystal on the front face of an ECAL module. The central hole in the plot corresponds to the beam hole.  }
\end{figure} 

\begin{figure} [b]
\centering
\begin{tabular}{c} 
$\vcenter{\hbox{\includegraphics[width=0.3\linewidth]{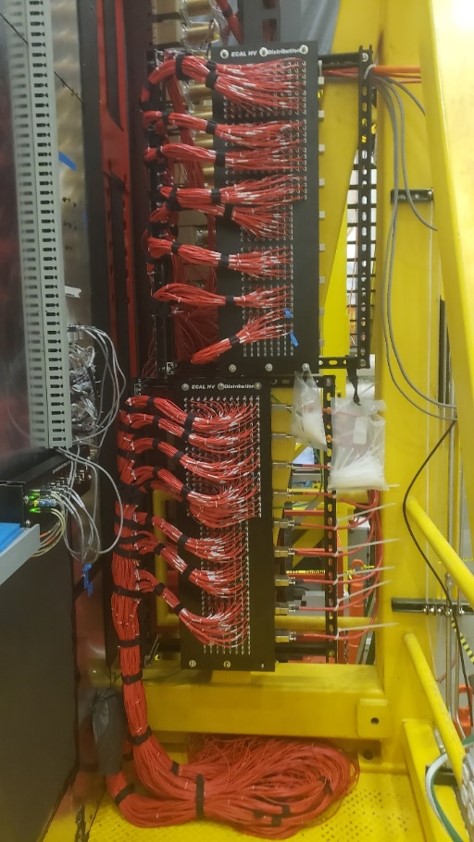}}}$
\hspace{0.25in}
$\vcenter{\hbox{\includegraphics[width=0.55\linewidth]{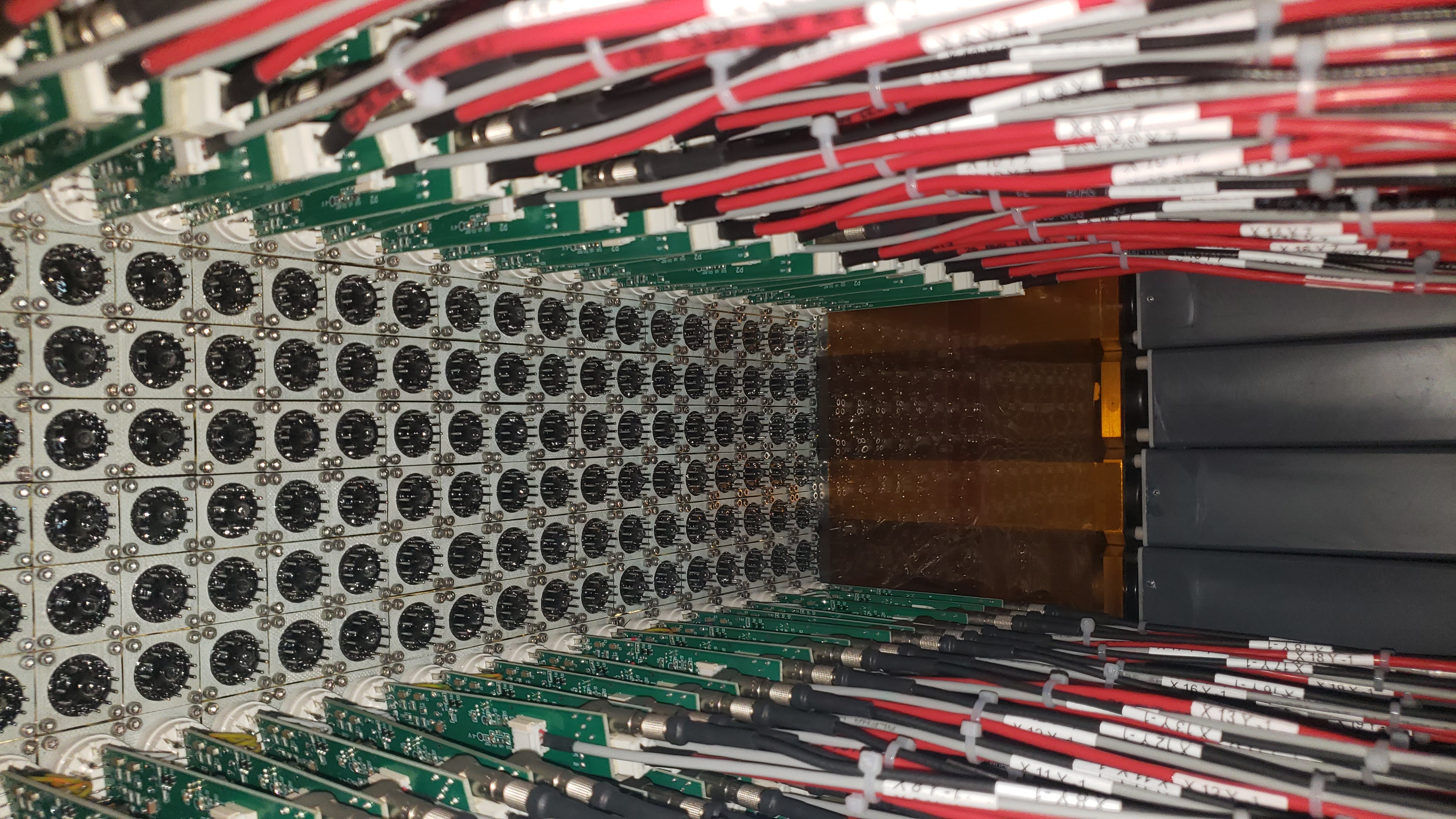}}}$
\end{tabular}
\centering
\vspace{0.2in}
\caption[example] 
{ \label{fig:ecal_divider} 
Two high voltage distribution boards installed near the detector dark box (left). Connecting active bases to the PMT sockets (right). Each divider is connected to three cables: signal, high voltage, and low voltage.}
\end{figure} 

Monitoring of the detector performance during the experiment is performed using a light monitoring system (LMS). The LMS generates light flashes, that are recorded during data taking and used to monitor the stability of the signal response over time. The light source consists of 25 blue LEDs, positioned inside an integrating sphere to ensure uniform light mixing. The mixed light is delivered to the face of each ECAL module via  $500\;\mu m$-thick optical fibers. Each fiber is glued to the face of the crystal using UV-curable glue. The installation of the optical fibers is shown in Fig.~\ref{fig:install_fibers}. The central hole in the plot corresponds to the beam hole. The ECAL modules surrounding this hole are covered by a tungsten absorber, which is used to protect them from the high-rate electromagnetic background.

\begin{figure} [b]
\begin{center}
\begin{tabular}{c} 
\includegraphics[height=8cm]{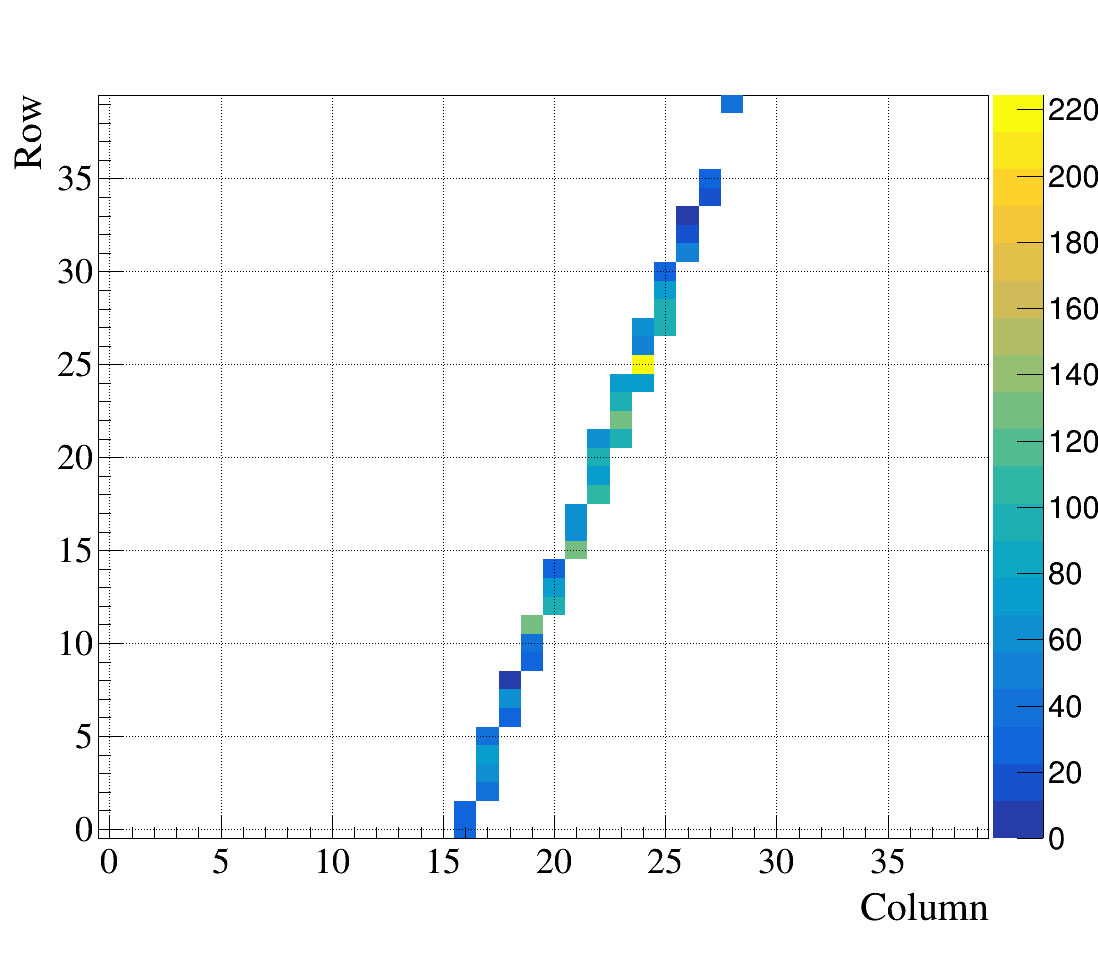}
\end{tabular}
\end{center}
\caption[example] 
{ \label{fig:cosmic_track} 
An example of the cosmic ray track reconstructed in the ECAL.}
\end{figure}

The installation of the ECAL cables was carried out in several steps. Signal, high-voltage (HV) and low-voltage (LV) cables (each approximately 50 feet long) were first routed from the electronics modules to distribution boards mounted on the frame of the detector dark room. Inside the dark room, short cables (up to 16 inches in length) were used to connect PMT active bases to the distribution boards. Three individual cables, signal, HV, and LV,  were installed and connected to each divider. Outside the dark room, the 1596 signal cables were connected to one hundred 16-channel flash ADCs modules positioned in 7 VXS crates. Thirty six high-voltage cables connect 36 48-channel CAEN 7030N modules placed in a CAEN HV SY4527 mainframe. Low voltage is supplied by two MPOD modules positioned in the Wiener crate, and distributed to the distribution boards via four cables. In total, approximately 6,500 cables were installed for the ECAL detector. Two high voltage distribution boards with the connected cables are presented in the left plot of Fig.~\ref{fig:ecal_divider}.

In the final stage of the detector installation, the cables were connected to the PMT active bases, which were then attached to the PMT sockets. The divider installation process is shown in the right panel of Fig.~\ref{fig:ecal_divider}. The cables were securely fastened to the vertical support chains to provide strain relief. To ensure proper PMT connections and cabling, we applied high voltage after installing the dividers on each detector layer and verified the module response using the LMS. These tests appeared to be critical, as the cable density limited access to the dividers after installation.

\begin{figure} [ht]
\begin{center}
\begin{tabular}{c} 
\includegraphics[height=5.5cm]{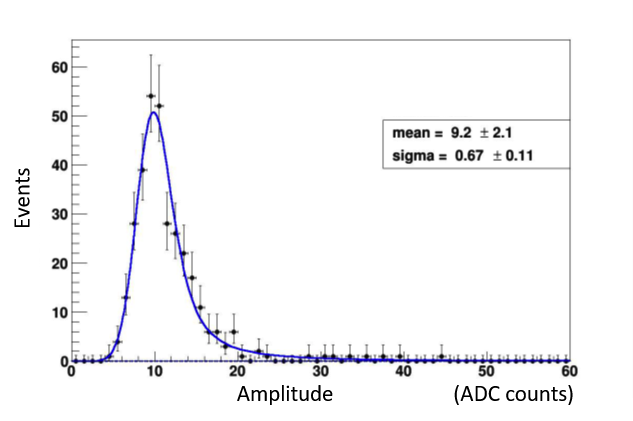}
\includegraphics[height=5.5cm]{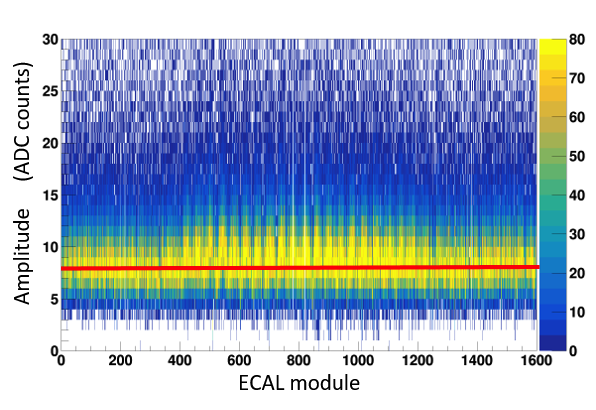}
\end{tabular}
\end{center}
\caption[example] 
{ \label{fig:cosmic_calib} 
An example of the fit to the distribution of flash ADC amplitudes induced by cosmic ray particles in the ECAL module (left). ECAL ADC amplitudes in the ECAL modules (right).}
\end{figure} 

\section{ECAL COMMISSIONING}
After installing the detector modules and connecting the cables, we proceeded with the initial commissioning of the detector using the light monitoring system and cosmic rays. The LMS was used to identify any potential issues with the detector following installation. The ECAL was successfully integrated into the GlueX detector’s data acquisition and trigger systems~\cite{trigger}. The right plot on Fig.~\ref{fig:testing_dividers} shows a flash ADC signal waveform induced by the LMS in an ECAL module. The LMS was also used to study the dependence of the signal pulse amplitude on the high voltage applied to each module. This dependence was used for subsequent voltage adjustment on individual modules. 

The cosmic ray particles were used to set up the initial high voltages for each ECAL module. Relativistic cosmic muons that penetrate the modules deposit a minimum amount of energy through ionization and are therefore referred to as minimum ionizing particles (MIPs). For a given path length, the energy response should be the same for all detector modules. Therefore, the high voltage of each ECAL module can be adjusted to equalize the signal pulse amplitudes produced by MIPs. This adjustment is particularly important for integrating the ECAL into the GlueX trigger, which assumes a uniform energy response across all modules for photons of the same energy. To detect cosmic ray particles, we installed two scintillating paddles, one above and one below the calorimeter. A coincidence of signals in these paddles was used to generate a cosmic ray trigger, which initiated the readout of the ECAL hits. An example of a cosmic ray track reconstructed in the ECAL is shown in Fig.~\ref{fig:cosmic_track}. The left plot of Fig.~\ref{fig:cosmic_calib} shows a typical distribution of flash ADC signal pulse amplitudes in the ECAL modules produced by cosmic ray muons. This distribution was fitted with a Landau function to determine the average flash ADC amplitude, corresponding to the response to a MIP. The high voltages on the ECAL modules were then adjusted to equalize these amplitudes and to set the average MIP response to 8 flash ADC counts. The resulting flash ADC amplitudes for all ECAL modules after adjusting HVs are shown in the right plot of Fig.~\ref{fig:cosmic_calib}. The PMT gain calibration was subsequently refined using beam data by reconstructing Compton scattering events and analyzing $\pi^0$ mesons. These procedures will be described in future publications.

\section{SUMMARY}
The Jefferson Lab Eta Factory experiment required upgrading the inner part of the forward lead glass calorimeter of the GlueX detector with high-resolution high-granularity PbWO$_4$ scintillating crystals forming the ECAL. Construction of the ECAL began in 2022. A total of 1596 modules were assembled, tested, and successfully installed on the GlueX detector. The ECAL was fully integrated with the GlueX infrastructure, including the readout and voltage supply electronics, as well as the data acquisition and trigger systems. A light monitoring system was installed and employed for the initial testing of the ECAL modules. Commissioning of the detector was subsequently performed using cosmic rays, which allowed setting initial high voltages on the detector. The JEF experiment began data taking in parallel with the GlueX-II experiment in April 2025. The ECAL is currently the largest PbWO$_4$-based calorimeter at Jefferson Lab.

\acknowledgments 
 This work was supported by the Department of Energy, USA. Jefferson Science Associates, LLC operated Thomas Jefferson National Accelerator Facility for the United States Department of Energy under contract DE-AC05-06OR23177. We thank the Jefferson Lab Physics Division groups and members of the participating universities for their valuable assistance with the ECAL fabrication and the installation of the detector in the experimental hall.

\bibliography{report} 
\bibliographystyle{spiebib} 

\end{document}